\documentclass[aps,prl,reprint,groupedaddress]{revtex4-1}
\usepackage[T1]{fontenc}
\usepackage[ansinew]{inputenc}
\usepackage{booktabs}
\usepackage{lmodern} %Type1-font for non-english texts and characters

\usepackage{microtype}

\usepackage{graphicx}
\usepackage[breaklinks]{hyperref}

\newcommand{\ghz}{\ensuremath{\, \mathrm{GHz}}}

\begin{document}

%\preprint{}

\title{Phase modulation and amplitude modulation interconversion for magnonic circuits}


\author{Calvin J. Tock}
%\thanks{}
\affiliation{Department of Physics, University of Oxford, Clarendon Laboratory, Parks Road, OX1 3PU}

\author{John F. Gregg}
%\thanks{}
\affiliation{Department of Physics, University of Oxford, Clarendon Laboratory, Parks Road, OX1 3PU}


\date{\today}

\begin{abstract}
Circuit elements within magnonic computers generally operate using signals that are either amplitude or phase modulated (AM or PM). We propose a simple all-magnon circuit element capable of converting between these two types of encoding. We highlight the potential of our technique to augment existing devices and propose a novel schematic for an `equality' gate and XNOR gate.
\end{abstract}

%\pacs{}
%\keywords{}

\maketitle

%% References should be done using the \cite, \ref, and \label commands
%\section{}
%\subsection{}
%\subsubsection{}

%\section{Introduction}

Magnon computing is an emerging field of study which aims to advance information processing beyond current CMOS technology \cite{Chumak2015,Csaba2014,Csaba2017,Toedt2016,Lenk2011,Khitun2013,Sadovnikov2018,Cornelissen2018,Nikitov2015}. The limits of Moore's law are rapidly being approached \cite{Waldrop2016}, and a new computing paradigm is required to help stave off the impending computational crisis caused by the overheating of progressively smaller silicon-based transistors. Magnonics is one such technology. Magnonic computers utilize spin waves, whose quanta are called magnons, to perform complex computational processes using a fraction of the energy and space required by conventional circuitry and with potentially enormous clock speeds \cite{Chumak2015}. With its current trajectory, magnonics looks to augment this conventional circuitry with specialist designed-for-purpose micro-circuits that can outperform standard devices. Several key elements of these circuits have so far been developed \cite{Chumak2014,Chumak2010,doi:10.1063/1.4898042,Chumak2017,Chumak2015}, some of which are discussed further below.

%Magnon computing is an emerging field of study which aims to advance information processing beyond current CMOS technology. The limits of Moore's law are rapidly being approached \cite{Waldrop2016}, and a new computing paradigm is required to help stave off the impending computational heat death. Magnonics is one such technology. REPHRASE THIS. Magnonic computers utilize spin waves, whose quanta are called magnons, to perform complex computational processes using a fraction of the energy and space required by conventional circuitry and with potentially enormous clock speeds \cite{Chumak2015}. With its current trajectory, magnonics looks to augment this conventional circuitry with specialist designed-for-purpose micro-circuits that can outperform standard devices. Several key elements of these circuits have so far been developed \cite{Chumak2014,Chumak2010,doi:10.1063/1.4898042,Chumak2017,Chumak2015}, some of which are discussed further below.

One aspect of these device elements whose significance is often neglected is the method by which data is encoded in the magnonic signal. For some devices, such as the majority gate \cite{doi:10.1063/1.4898042}, the digital data is encoded in the phase of the magnons, whereas for the all-magnon transistor \cite{Chumak2014}, the amplitude is the relevant quantity. In a real magnonic computer where both types of components will be used, it will be necessary to convert between phase modulated and amplitude modulated signals. Some conversion methods have already been proposed \cite{Bracher2016}, however, these require external electronics. In this paper, we propose an all-magnon based method for PM to AM (P2A) and AM to PM (A2P) conversion.

As we will be discussing the use of two different binary data encoding methods, it is convenient to shift away from the nomenclature of purely 1s and 0s, and move to a system that provides the flexibility to discuss both modulation techniques simultaneously. The phase and amplitude of a magnon with respect to a reference source can be recorded in the form $\zeta(\textrm{R},\theta,\phi)=\textrm{R}\textrm{e}^{i(\theta+\phi)}$, where $\theta$ encodes the phase modulated binary state, which for one such implementation takes the values of 0 or $\pi$, and $\phi$ is a controllable phase offset. R represents the relative magnitude of the signal. If $\phi$ is set to zero, the phase encoded binary states can be written as $\zeta(\textrm{R},\{0,\pi\},0)=\{-\textrm{R},\textrm{R}\}$. For amplitude modulation, the signal is either zero or non-zero, and can take any phase, and can be written as $\zeta(\{0,\textrm{R}\},\theta,\phi)=\{0, \textrm{Re}^{i(\theta+\phi)}\}$. By combining phase and amplitude modulation and setting the amplitude to one, a simple example of a three-level mixed AM/PM system with $\zeta(\{0,1\},\{0,\pi\},0)=\{-1,0,1\}$ is obtained, which is a useful starting point for the discussion below.

\begin{figure*}
	\centering
		\includegraphics[width=0.9\textwidth]{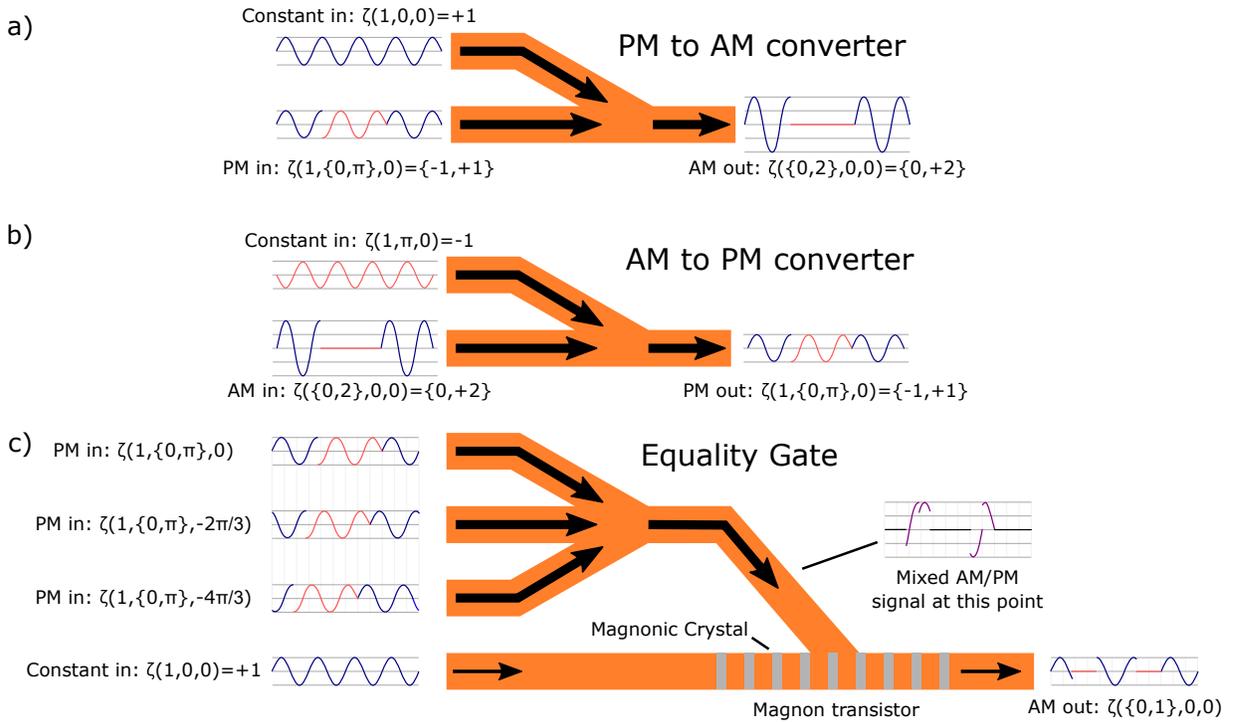}
	\caption{(Color online) a) Schematic of the Phase to Amplitude modulation (P2A) converter with example waveforms. Binary 1's and 0's are indicated in red (light gray) and blue (dark gray). b) The Amplitude to Phase modulation (A2P) converter, again with example waveforms. c) Schematic of the Equality Gate, which takes a similar form to the majority gate \cite{doi:10.1063/1.4898042}, followed by the all-magnon transistor \cite{Chumak2014}. The three input channels are each separated by $\frac{2\pi}{3}$, and the output is 1 when all three inputs are in the same binary state.}
	\label{fig:AMPMEquality}
\end{figure*}

A method of converting between phase and amplitude modulated signals is proposed as follows: both types of converters work by interfering two magnon channels at a junction, with the incoming data signal entering from one of the input branches (Figure \ref{fig:AMPMEquality} a and b). The other input comes from a pure tone continuous magnon source with a specific phase and amplitude. 

\begin{figure}[htbp]
	\centering
		\includegraphics[width=0.45\textwidth]{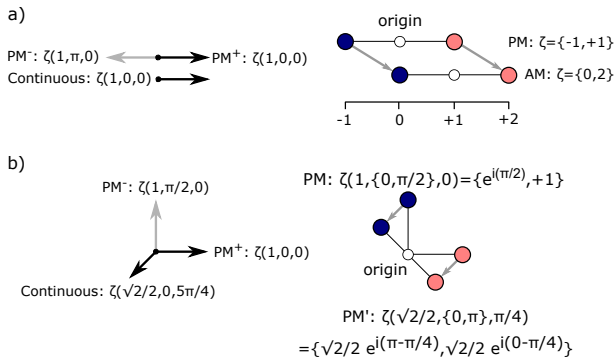}
	\caption{(Color online) a) Constellation plot for the P2A converter. The $\zeta=\{-1,1\}$ PM signal values shift to a $\zeta=\{0,2\}$ AM signal with the application of the continuous wave. b) Example of a Phase to Phase modulation (P2P) transformation, in which a PM signal with $\zeta(1,\{0,\frac{\pi}{2}\},0)$, (which has a $\Delta\theta=\frac{\pi}{2}$) is converted to a new PM' signal with $\zeta(\frac{\sqrt{2}}{2},\{0,\pi\},\frac{-\pi}{4})$, which has a $\Delta\theta=\pi$.}
	\label{fig:PM2PM}
\end{figure}

For the phase to amplitude conversion (P2A) where the signal input takes the values of $\zeta=\{-1,1\}$, the continuous-wave input is set to $\zeta=+1$. When the two pulse trains combine and interfere (constructively for $+1$ and destructively for $-1$), the output will be an amplitude modulated signal with $\zeta={0,2}$, which can be attenuated if necessary (Figure \ref{fig:PM2PM}a). The reverse process (A2P) works by setting the continuous input to -0.5. In this case, an AM signal input of $\zeta=\{0,1\}$ results in a phase modulated output of $\zeta=\{-0.5,0.5\}$, which can be amplified if required. 

\begin{figure}[htbp]
	\centering
		\includegraphics[width=0.45\textwidth]{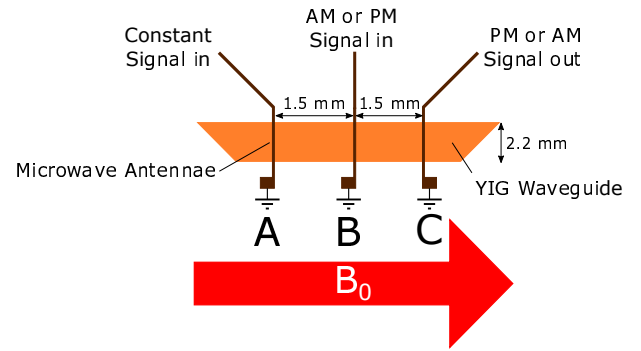}
	\caption{The experimental arrangement used to test the A2P and P2A converters.}
	\label{fig:YIGexperiment}
\end{figure}

This concept has been experimentally verified using a linear YIG waveguide in the backward volume mode in the presence of three equally spaced antennae (Figure \ref{fig:YIGexperiment}). Pulsed AM or PM signals of $3.2\ghz$ microwaves were inserted into antenna B, while the pure-tone signal was applied to antenna A. The electrical signal collected by antenna C was homodyned with a reference signal and then passed through a low-pass filter to remove the $2f$ component. The trace of the (modulated) DC component was then recorded on an oscilloscope. When the amplitude of the pure-tone signal was set correctly, both the A2P and P2A arrangements functioned as expected. 

By controlling the phase and amplitude of the continuous pure-tone magnon source, additional functionality is available. For example, this method can be used to convert between different types of phase modulated signals, such as the PM signals used in \cite{Bracher2016} which use a modulation of $\theta = \{0,\frac{\pi}{2}\}$. By adding a continuous signal with $\zeta(\frac{\sqrt{2}}{2},0,\frac{5\pi}{4})$, a new PM signal with $\Delta\theta = \pi$ can be obtained (Figure \ref{fig:PM2PM}b). This technique can also be used as an AM NOT gate by simply adding a wave with $\zeta=-1$ to an AM signal with $\zeta=\{0,1\}$. 

By using these conversion techniques, it is possible to interface PM driven devices such as the majority gate with AM driven devices like the magnon transistor or amplifier \cite{Chumak2014}. By considering devices with mixed AM/PM encodings, it is suggested that this technique opens new possibilities for magnonic circuit design, such as the equality gate.

\begin{figure*}
	\centering
		\includegraphics[width=0.7\textwidth]{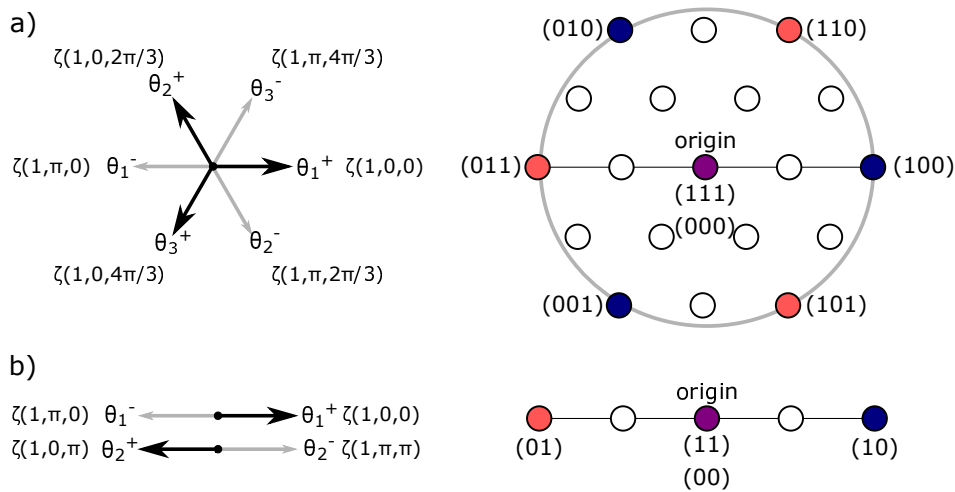}
	\caption{a) Constellation plot of the 8 possible intermediary outcomes of the equality gate, prior to rectification in the magnon transistor. The two `equality' states lie at the origin, whilst the other states lie on a circle of constant radius (amplitude). b) Constellation plot of an XNOR gate (the two channel analogue of the equality gate), in which two $\pi$-separated PM signals interfere at a junction. The same method of rectification (the magnon transistor) can again be used to extract the digital information.}
	\label{fig:constelationequality}
\end{figure*}

The equality gate has the same appearance as the majority gate followed by an all-magnon transistor (Figure \ref{fig:AMPMEquality}c), and uses PM inputs with $\theta = \{0,\pi\}$. The key difference in this device is that the three input channels are each separated by a fixed phase of $\Delta\phi=\frac{2\pi}{3}$.  The eight possible output states from the majority-like section can be calculated and plotted in a constellation plot on the Argand diagram. (Figure \ref{fig:constelationequality}a).
 
In Figure \ref{fig:constelationequality}a, the two states for which all three inputs take equal logic values (111 , 000) lie on the origin, whereas the other six states lie on a circle with a constant radius. By extracting the AM information from this mixed AM/PM output state, this device is able to function as an `equality' gate. This operation is performed by the subsequent all-magnon transistor \cite{Chumak2014}.

\begin{table}[btp]
\centering
\label{tab:truthtable}
\begin{tabular}{ccc|c}
\toprule
%\multicolumn{3}{l}{Inputs} & Output \\ \hline
Input A       &Input B       &Input C      &Output      \\ \hline
0       & 0       & 0      & 1      \\
0       & 0       & 1      & 0      \\
0       & 1       & 0      & 0      \\
0       & 1       & 1      & 0      \\
1       & 0       & 0      & 0      \\
1       & 0       & 1      & 0      \\
1       & 1       & 0      & 0      \\
1       & 1       & 1      & 1      \\
\bottomrule
\end{tabular}
\caption{Truth table for the equality gate.}
\end{table}

As the transistor only relies on the presence of the gating magnons within the magnonic crystal, the phase information of the input signal is irrelevant. As the 111 and 000 states have no amplitude and thus no magnons, the transistor will be open and produce an output value of $\zeta=+1$. The other six states will close the transistor (by the same amount, as they all have the same amplitude) and produce an output of $\zeta=0$. These values represent AM encoded logic which agrees with the truth table for the equality gate shown in table \ref{tab:truthtable}.

It is worthy of note that this equality gate arrangement of multiple converging phase-separated magnon signals followed by an all-magnon transistor also functions as an XNOR gate when only two inputs separated by $\pi$ radians are used (Figure \ref{fig:constelationequality}b). We also highlight that by combining an equality gate with a majority, NOT, and XOR gate, it is possible to construct a full adder circuit.

In conclusion, we have presented a simple method for converting between amplitude and phase modulated magnonic signals using wave interference at a waveguide junction. We have shown that this technique can be used to convert between different types of phase modulation, as well as provide the basis for other simple logic operations such as an AM NOT gate. We have suggested that magnonic circuit elements can be designed that utilize both AM and PM encoding simultaneously, and have proposed the schematic of two such devices: the equality gate and an XNOR gate. We believe these new circuit elements complement the plethora of already existing magnonic components, and provides a missing piece of the puzzle that enables the sequentialization of magnonic devices.

\bibliography{AMPM2bib}

\end{document}